\documentclass{PoS}
\usepackage{amssymb}
\usepackage{amsmath}
\usepackage{graphicx}
\usepackage{epsfig}
\usepackage{graphics}

\newcommand{\be}{\begin{eqnarray}}
\newcommand{\ee}{\end{eqnarray}}
\title{Gravitational waves from precessing engines in GRBs}

\ShortTitle{Gravitational waves from precessing engines in GRBs}

\author{{Gustavo E. Romero}\\
        Instituto Argentino de Radioastronom\'{\i}a (IAR), CCT
La Plata  (CONICET), Argentina, and Facultad de Ciencias Astron\'omicas y Geof\'{\i}sicas,
Universidad Nacional de La Plata (UNLP), Paseo del Bosque s/n, 1900, La Plata, Argentina \\
        E-mail: \email{romero@fcaglp.unlp.edu.ar}}

\author{{Mat\'{\i}as M. Reynoso}\\
Instituto de
Investigaciones F\'{\i}sicas de Mar del Plata (CONICET - UNMdP), Universidad Nacional de
Mar del Plata, Dean Funes 3350, (7600) Mar del Plata, Argentina\\
        E-mail: \email{mreynoso@mdp.edu.ar}}

\author{{Hugo R. Christiansen}\\
State University of Cear\'a, Physics Dept., Av. Paranjana 1700, 60740-000 Fortaleza - CE, Brazil;
 Universidade Estadual Vale do Acara, Av. da Universidade 850, 62040-370 Sobral - CE, Brazil  \\
        E-mail: \email{hugochristiansen@yahoo.com.br}}

\abstract{In the framework of a transient accretion disk at the core
of a gamma-ray burst we compute possible periods of Lense-Thirring precession.
Next, we evaluate the putative gravitational waves associated with such dynamical setup.
\ Assuming a characteristic time-profile for the gamma-ray emission
of a disk-jet system, we obtain light-curves presenting a time microstructure
similar to that reported in some GRB events.
After adjustment of the parameters out of two specific GRBs
we evaluate the detectability of the gravitational waves produced by
the precession of this accretion disk.
As a conclusion, our analysis shows that some GRBs are likely to be probed with Advanced LIGO.}

\FullConference{25th Texas Symposium on Relativistic Astrophysics - TEXAS 2010\\
		December 06-10, 2010\\
		Heidelberg, Germany}

\begin{document}

\section{Introduction}
It is widely thought that gamma-ray bursts (GRBs) are stellar events produced out of
a central object that consists of a massive black hole with
a transient, hot and dense accretion disk. Such an engine might result
from a failed supernova or collapsar (e.g.
\cite{Woosely 1993}) or the merging of two compact objects (e.g. \cite{Mochkovitch
et al. 1993}). Apparently, the accretion of matter onto the black hole is the energy source of the system.
Part of this energy is ejected through an ephemeral relativistic bipolar jet,
normal to the midplane of the disk,  radiating an
extremely intense burst  of gamma-rays. The duration of these
 gamma-ray pulses ranges from milliseconds up to minutes. This picture is expected to
underly most of the observed GRB light-curves.

The usual explanation for the temporal structure of such light curves
relays on the formation of shock waves that convert bulk kinetic energy into
relativistic energy of the particles in the jets. Charged particles then cool by
synchrotron and inverse Compton emission. The shocks can be either
internal to the jet
and produced by  colliding shells with different Lorentz factors
(e.g. \cite{KobayashiPiranSari 1997, DaigneMochkovitch 1998,
GuettaSpadaWaxman 2001}) or the result of interactions with
the external medium (e.g. \cite{HeinzBegelman 1999}).

Nevertheless, some peculiar GRB light curves
are difficult to explain  by means of shocks only, particularly
 those having slow rises and fast decays
(e.g. \cite{Romero et al. 1999}). In this respect, it has been suggested that the
precession of the jet might play an important role in the formation of its time
microstructure and this could be valid in both long and short gamma-ray bursts (e.g.
\cite{Blackman et al. 1996, Portegies et al. 1999, Fargion 1999, Reynoso
et al. 2006}). In \cite{Reynoso et al. 2006} a model for precessing jets
has been based on spin-induced precession of a neutrino-cooled massive disk. We will
use this model to obtain information about the inner system from the GRB light-curves.

Our goal here is the study of another expected byproduct of disk
precession, namely the production of gravitational waves.
Strong gravitational radiation is likely in several situations:
when the gravitational collapse originating the burst is non-spherical,
in the presence of strong inhomogeneities in the accretion disk, or, for
 short GRBs, as a result of the spiral merging of
compact objects (e.g. \cite{Mineshinge et al. 2002,  KobayashiMeszaros 2003}).

In the next sections we will present a model for the production
of a gamma-ray light curve by a central engine consisting of a massive
black hole, a transient accretion disk  and a jet. Considering Lense-Thirring precession
we will then compute gravitational waves emitted by such a source and
finally discuss their possible detection.

\section{Accretion disk and spin-induced precession}

In any model for a transient accretion disk formed
after massive collapse or merging of compact objects, it is reasonable to consider it
to be initially misaligned with the resulting black hole.
As first noticed by Lense and Thirring in the general case \cite{LenseThirring 1918},
in the central engine of a GRB this misalignment will
cause disk precession. This phenomenon is originated in the dragging suffered by the
inertial frames near a rapidly spinning black hole.

When the Mach number inside the disk is below 5, which seems to be the case in most GRBs,
it is likely that the disk precesses approximately like a rigid body, i.e.,
it presents no warping \cite{NelsonPapaloizou 2000}.
Of course, the precession of the disk should lead to the
precession of the jets, yielding a likely source of temporal micro-variability
in the gamma-ray signal \cite{Reynoso et al. 2006}.

The typical disk accretion rate, ranging from $\sim
0.1$ to  $10 M_\odot \; \rm s^{-1}$, is expected to significantly vary
 in the outer part of the disk. In contrast, for the inner disk a constant accretion
rate should be a valid approximation (e.g. \cite{Popham 1999, Di Matteo et
al. 2005}).
The conservation of mass, falling with a velocity $v_r\simeq r
\sqrt{GM_{\rm bh}r^{-3}}$ at a distance $r$ from the black hole
axis, is given by
 \be
 \dot{M}= -2\pi r v_r \Sigma(r)
 \ee
where $\Sigma(r)= 2 \rho(r) H(r)$ is the surface density, $H(r)$
is the disk half-thickness, and $\rho(r)$ is the mass density of
the disk. Now, conservation of angular momentum and energy can be
used to numerically compute $\Sigma(r)$ and $H(r)$,
considering that the heat generated by friction can be balanced by
advection and neutrino emission \cite{Reynoso et al. 2006}.

Without nutation, the precession period of the
disk $\tau_\mathrm{p}$ can be related to its surface density by
 \cite{LiuMelia 2002, Reynoso et al. 2006}
 \be
  \tau_p=\int_{0}^{2\pi} \frac{L_{\rm d}}{T_{\rm d}}\sin \theta d\phi = 2
  \pi \sin \theta \frac{L_{\rm d}}{T_{\rm d}},
 \ee
where the values of the disk angular momentum $L_{\rm d}$ and applied
precessional torque $T_{\rm d}$  are
 \be L_{\rm d} &=& 2
 \pi\int_{R_\mathrm{ms}}^{R_\mathrm{out}} \Sigma(r) \Omega_\mathrm{k}(r) r^3 \ dr, \\
 T_{\rm d}&=&4\pi^2 \sin{\theta}
 \int_{R_\mathrm{ms}}^{R_\mathrm{out}} \Sigma(r) \Omega_\mathrm{k}(r) \nu_{p,\theta}(r)
 r^3 \ dr.
 \ee
The relativistic Keplerian angular velocity reads
  \be \Omega_\mathrm{k}(r)=\frac{c^3}{GM_\mathrm{bh}} \left[
\left( \frac{r}{R_\mathrm{g}} \right)^{3/2}+a_*  \right]^{-1}, \ee
where $a_*$ is the spin parameter,
$R_\mathrm{g}=GM_\mathrm{bh}/c^2$ the gravitational radius, and
\be \nu_{p,\theta}=\frac{\Omega_k(R)}{2\pi} \left[1 - \sqrt{1\mp
4a_*\left(\frac{R_\mathrm{g}}{r}\right)^{1/2}+3a_*^2
\left(\frac{R_{\rm g}}{r}\right)^2}\right] \ee is the nodal
frequency resulting from the perturbation of a circular orbit in the Kerr
metric \cite{Kato 1990}. Since the action of viscous torques
leads to the alignment of the
very inner part of the accretion disk with the black hole equator
plane \cite{BardeenPetterson 1975}, the precessing part of the disk ends at an
outer radius $R_{\mathrm{out}}$, extending from an inner radius
$R_\mathrm{ms}= \xi_{\rm ms}R_\mathrm{g}$, where \be
\xi_\mathrm{ms}=3 +A_2 \mp [(3-A_1)(3+A_1+2A_2)]^{1/2}, \ee with
\be A_1=1+(1-a_*^2)^{1/3}[(1+a_*)^{1/3}+(1-a_*)^{1/3}], \ee and
\be A_2=(3a_*^2+A_1^2)^{1/2}. \ee The minus sign in $\xi_{\rm ms}$
corresponds to prograde motion ($a_*>0$), whereas the plus sign
corresponds to retrograde motion ($a_*<0$).

\section{Gravitational waves from Lense-Thirring precession}

The gamma-ray luminosity produced by a relativistic jet
depends on the observation angle $\psi$ as given by \cite{Portegies et al. 1999}
 \be
 L(\psi)= \frac{27}{4}\left[e^{-0.6 \Psi(x)}-\frac{8}{9}\right]  \left[ e^{-0.3 \Psi(x)}- e^{-6.3 \Psi(x)}\right]\frac{d\Psi(x)}{dx}. \label{Lum}
 \ee
Here, $\psi$ is the angle between the jet and the observer, $x=10 \sin\psi$ and $\Psi(x)= \frac{1}{6}\ln(1+4x^2)$.
The intrinsic time dependence of the light signal, on the other hand, might be characterized by a FRED (Fast Rise and Exponential Decay) function,
\be
I(t) = N_I \left( 1-e^{-\frac{t}{\tau_{\rm rise}}} \right) \left\{ \frac{\pi}{2}-
\tan^{-1} \left[ \frac{t-\tau_{\rm plat}}{\tau_{\rm dec}} \right]\right\}, \label{Idt}
\ee
where $N_I$ is a normalization constant such that the maximum of
the signal corresponds to unity. The initial
rise, plateau, and decay timescales are $\tau_{\rm rise}$, $\tau_{\rm
plat}$, and $\tau_{\rm dec}$  respectively.

\begin{figure}[h]
  \centering
\includegraphics[trim=0 0 0 0,clip]{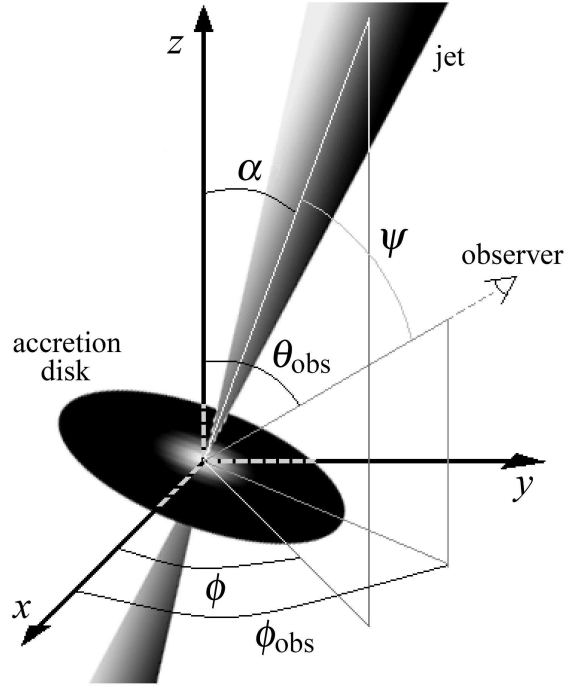}
  \caption{Pictorial scheme of a GRB engine.}
  \label{FigGRBscheme}
\end{figure}

In order to make our predictions we shall consider two specific
GRBs, the \textit{short} GRB 990720 and the \textit{long}  GRB 990712.
For these events, we use $F(t)=I(t)\times L(\psi(t))$ to reproduce the observed light curves. Here, the angle between the jet and the observer is time-dependent because the azimuthal angle of the jet is $\phi(t)= 2\pi (t/\tau_{\rm prec})$ as a result of the disk/jet precession (see Fig.\ref{FigGRBscheme}). Both angles are related by
\be
\cos{\psi}= \hat{r}_{\rm jet}\cdot \hat{r}_{\rm obs}\
       = \cos{\phi}\sin{ \alpha } \sin{\theta_{\rm obs}} + \cos{\alpha}\cos{\theta_{\rm obs}}.
\ee
Taking this into account, we can obtain the relevant timescales and precession period by the procedure
above described. In both cases, the observer is located at $\theta_{\rm obs}=2^\circ$ with respect to the $z$-axis perpendicular to the black hole equator, and we set $\phi_{\rm obs}$ at $\phi= 0$ (see Fig.\ref{FigGRBscheme}).

\begin{figure}[h]
  \centering
\includegraphics[trim=0 0 0 0,clip, width= .6 \linewidth]{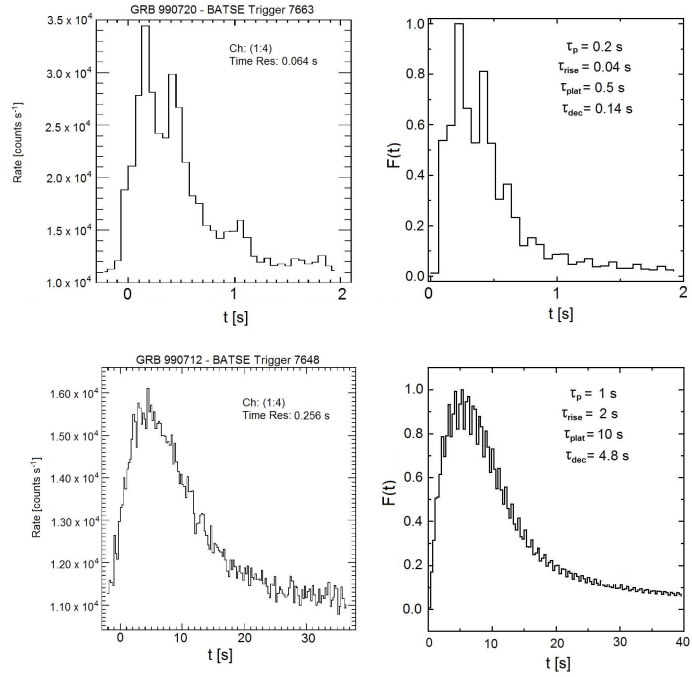}
  \caption{(up) GRB 990720 light curve, experimental (left panel) and model obtained (right panel).  (down) GRB 990712 light curve, experimental (left panel) and model obtained (right panel). }
  \label{FigGRBlong}
\end{figure}

\subsection{Emission of gravitational waves }

Axis-symmetric bodies (i.e. with inertial moments $I_1=I_2$) in
precession emit gravitational waves with an amplitude given by
\cite{ZimmermannSzedenits 1979, Maggiore 2008}
 \be
 h_{\rm prec}(t)= h_+(t)+h_\times(t), \label{hprec}
 \ee
where
 \be
 h_{+}(t)= F_{+,1} \cos{\Omega t} + F_{+,2} \cos{2\Omega t},\ \ \
 h_{\times}(t)= F_{\times,1} \sin{\Omega t} + F_{\times,2} \sin{2\Omega t},\label{h+x}
 \ee
%
with
 \be
  F_{+,1}&=& h'_0 \sin{2\alpha} \sin{\iota} \cos{\iota},\ \ \
  F_{+,2}= 2 h'_0 \sin^2{\alpha} (1+\cos^2{\iota})\nonumber \\
  F_{\times,1}&=& h'_0 \sin{2\alpha} \sin{\iota},\ \ \ \ \ \
  F_{\times,2}= 4 h'_0 \sin^2{\alpha} \cos{\iota},\nonumber
 \ee
and
$
 h'_0= -\frac{G}{c^4}\ {(I_3-I_1)\Omega^2}/{d}.
$
Here, $\alpha$ is the angle between the angular momentum of the
disk and that of the black hole (see Fig.\ref{FigGRBscheme}), $\iota$ is the angle between the
$z$-axis of the detector and the signal direction of arrival (line of sight), and
$d$ is the distance to the radiating body. The principal
moments of inertia are
$
 I_3=\int_V (x^2+y^2) \rho(r), \
 I_1=\int_V (z^2+y^2) \rho(r).
$
The frequency of the gravitational waves (GW) are $f_1=\Omega/(2\pi)$ and
$f_2=2\Omega/(2\pi)$, which are related to the angular momentum of the body
by
$
 \Omega= {L}/{I_1}.
$

Since we are dealing with bursting events, the GW signal is expected to be significant for a brief
time $\tau_{\rm plat}$.
We therefore modulate the signal given by Eq. (\ref{hprec}) with a gaussian
 \be
 h(t)=h_{\rm prec}(t) \ e^{-\frac{t^2}{2\tau_{\rm plat}^2}}\label{hdet}
 \ee
as usually done to describe GW signals from bursting sources (e.g. \cite{Maggiore 2008, Acernese 2008}.
The angular frequencies that contribute to the waveform of Eq. (\ref{hdet}) can be better obtained looking at the Fourier transform of Eq.(\ref{hprec})
$\tilde{h}(\omega)= \tilde{h}_+(\omega) + \tilde{h}_\times(\omega) $, where
 \be
 \tilde{h}_+(\omega)= \frac{\tau_{\rm plat}}{2} \left[ F_{+,1}\left(e^{ \frac{-\tau^2_{\rm plat}}{2}(w+\Omega)^2 }+ e^{\frac{-\tau^2_{\rm plat}}{2}(w-\Omega)^2}\right)
 + F_{+,2}\left(e^{\frac{-\tau^2_{\rm plat}}{2}(w+2\Omega)^2 }+ e^{ \frac{-\tau^2_{\rm plat}}{2}(w-2\Omega)^2 }\right) \right]\nonumber
 \ee
 and
 \be
 \tilde{h}_\times(\omega)= \frac{i \tau_{\rm plat}}{2} \left[ F_{\times,1}\left(e^{ \frac{-\tau^2_{\rm plat}}{2}(\omega+\Omega)^2 }+ e^{\frac{-\tau^2_{\rm plat}}{2}(\omega-\Omega)^2}\right)
 +  F_{\times,2}\left(e^{\frac{-\tau^2_{\rm plat}}{2}(\omega+2\Omega)^2 }+ e^{ \frac{-\tau^2_{\rm plat}}{2}(\omega-2\Omega)^2 }\right) \right].\nonumber
  \ee
{From these expressions it is apparent that the main contributions to the signals
are close around $\Omega$ and $2\Omega$,
and the width of the interval is $\tau_{\rm plat}^{-1}$. Given the typical durations of GRBs, the frequency spread is narrow for all burst except for those with durations much shorter than $1$ s.
\begin{figure*}[]
  \centering
  \includegraphics[trim=15 200 0 18,clip, width=0.99\textwidth, keepaspectratio]{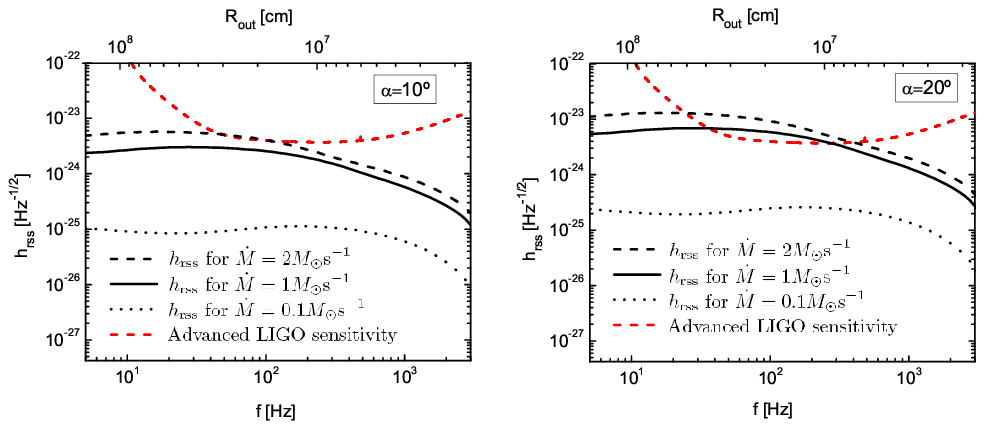}  
  \caption{Gravitational wave \textit{rss} amplitude for different accretion rates and Advanced LIGO sensitivity (red dashed line) as a function of the gravitational wave frequency, for $\alpha=10^\circ$ (left panel) and  $\alpha=20^\circ$ (right panel). The corresponding outer radius of the accretion disk is indicated in the upper horizontal axis.}
  \label{Fig_h0deRout}
\end{figure*}

\subsection{Numerical predictions of Lense-Thirring gravitational waves}

In order to judge detectability, we compute the root-sum-square amplitude
(e.g. \cite{Maggiore 2008, Acernese 2008}) using Eqs. (\ref{h+x})
 \be
 h_{\rm rss}(f)= \sqrt{\int_{-\infty}^{\infty}dt \, ( h_+^2+ h_\times^2) \, e^{-\frac{t^2}{\tau_{\rm plat}^2}}}.
 \ee
For illustration, we choose $\iota=45^\circ$ and  $\alpha=10^\circ, 20^\circ$.
{In Fig. \ref{Fig_h0deRout}
we plot the root-sum-square amplitude $h_{\rm rss}$,
as a function of one of the resonant frequencies, $f=\Omega/(2\pi)$}, and
also as a function of the outer radius of the inner precessing
disk. In the figure we include the expected sensitivity for
Advanced LIGO \cite{Shoemaker 2010}.
The parameters used for finding $\Sigma(r)$ and
$H(r)$ are $M_{\rm bh}= 3 M_\odot$, $a_*=0.1$, a viscosity
parameter $\alpha=0.1$, and the different mass loss rates are
$\dot{M}=\{0.1 M_\odot \;{\rm s}^{-1}, 1 M_\odot \; {\rm
s}^{-1},10 M_\odot \; {\rm s}^{-1}\}$. The distance to the GRB is
taken as $d= 100\; {\rm Mpc}$.

Fig. \ref{Fig_h0deRout} makes clear that there are better
chances of detection for accretion rates higher than $1M_\odot \;
{\rm s}^{-1}$ and outer radii between $10^{7}$ and $10^{8}$ cm.
Note that when the accretion rate is very high the disk may become advection
dominated rather than cooled by neutrino emission \cite{Liu et al.
2008} but this should not affect the dynamics in the gravitational field.
On the other hand, large accretion rates can be only supported in long
GRBs, so we conclude that there is only a good prospect for
detection of gravitational waves from precessing disks of nearby
($d<100$ Mpc) long events. Such events are likely related to
the death of very massive stars, so the host galaxies should have
active star forming regions.

Two low-luminosity \textit{long} GRBs (980425 and 060218) were already
observed at distances of $\sim$ 40 Mpc and $\sim$ 130 Mpc,
respectively \cite{CorsiMeszaros 2009}. The local rate of
long GRBs is estimated to be $\sim$ 200 Gpc$^{-3}$ yr$^{-1}$ (e.g.
\cite{Liang et al. 2007, Virgili et al. 2009}) and INTEGRAL has detected a
significant ratio of faint GRBs which are inferred to be local \cite{Foley at al. 2008}.
As a consequence, this work shows that
in the near future the detection of precessing disks
in the central engines of GRBs should be possible through their
gravitational emission \cite{RomeroReynosoChristiansen}.
The detection of one event of this class can be
used to test the Lense-Thirring effect in the strong field limit.

\noindent
{\textbf{Acknowledgements}:}
O.A. Sampayo is acknowledged for useful comments.
This research was supported by the Argentine Agencies CONICET and
ANPCyT through grants PIP 0078 and PICT-2007-00848 BID 1728/OC-AR. GER
acknowledges additional support from the Spanish MICINN
under grant AYA2010-21782-C03-01.

\end{document}